\documentclass[english,aps,prl,twocolumn]{revtex4}
\usepackage[T1]{fontenc}
\usepackage[latin9]{inputenc}
\usepackage{color}
\usepackage{units}
\usepackage{amssymb}
\usepackage{graphicx}
\usepackage{esint}

\makeatletter
\@ifundefined{textcolor}{}
{%
 \definecolor{BLACK}{gray}{0}
 \definecolor{WHITE}{gray}{1}
 \definecolor{RED}{rgb}{1,0,0}
 \definecolor{GREEN}{rgb}{0,1,0}
 \definecolor{BLUE}{rgb}{0,0,1}
 \definecolor{CYAN}{cmyk}{1,0,0,0}
 \definecolor{MAGENTA}{cmyk}{0,1,0,0}
 \definecolor{YELLOW}{cmyk}{0,0,1,0}
}

\usepackage{graphicx}
\usepackage{dcolumn}
\usepackage{bm}
\makeatletter

\newcommand{\Rmnum}[1]{\expandafter\@slowromancap\romannumeral #1@}
\makeatother

\makeatother

\usepackage{babel}
\begin{document}

\title{Probing long-range correlations in the Berezinskii-Kosterlitz-Thouless
fluctuation regime of ultra-thin NbN superconducting films using transport
noise measurements}

\author{R. Koushik$^{1}$, Siddhartha Kumar$^{1}$, Kazi Rafsanjani Amin$^{1}$,
Mintu Mondal$^{2}$, John Jesudasan$^{2}$, Aveek Bid$^{1}$, Pratap
Raychaudhuri$^{2}$ and Arindam Ghosh$^{1}$}

\address{$^{1}$ Department of Physics, Indian Institute of Science, Bangalore
560012, India}

\address{$^{2}$ Tata Institute of Fundamental Research, Homi Bhabha Road,
Colaba, Mumbai 400005, India}

\email{koushikr.in@gmail.com}

\begin{abstract}
We probe the presence of long-range correlations in phase fluctuations
by analyzing the higher-order spectrum of resistance fluctuations
in ultra-thin NbN superconducting films. The non-Gaussian component
of resistance fluctuations is found to be sensitive to film thickness
close to the transition, which allows us to distinguish between mean
field and Berezinskii-Kosterlitz-Thouless (BKT) type superconducting
transitions. The extent of non-Gaussianity was found to be bounded
by the BKT and mean field transition temperatures and depend strongly
on the roughness and structural inhomogeneity of the superconducting
films. Our experiment outlines a novel fluctuation-based kinetic probe
in detecting the nature of superconductivity in disordered low-dimensional
materials.
\end{abstract}
\maketitle
The transition from superconducting to normal state in two-dimensions
is known to occur via Berezinskii-Kosterlitz-Thouless (BKT) mechanism.
Traditionally, the signature of BKT transition is found by measuring
the superfluid density where a discontinuity is observed at the critical
temperature $T_{BKT}$~\cite{Berezinskii1972,Kosterlitz1973,Nelson1977}.
This method is limited in its scope of application for systems exhibiting
interfacial/buried superconductivity like oxide heterostructures due
to the inability to measure the thickness of the superconducting layer
accurately. Transport-based probes including discontinuity in the
power law behavior of I-V characteristics or change in the curvature
of magnetoresistance from convex to concave in the presence of a perpendicular
magnetic field~\cite{Halperin1979,MinnhagenPRB1981,KadinPRL1981,KadinPRB1983,SimonPRB1987,MinnhagenRMP}
have their limitations as real systems contain some degree of inhomogeneity
which smear out the signatures of BKT type behavior. With growing
interest in low-dimensional superconductivity, it is interesting to
develop new probes/techniques that are not only sensitive enough to
detect BKT transition but also compare its characteristics scales
with the mean field description.

The BKT transition, as exhibited by ultra-thin superconducting films,
is characterized by the unbinding of vortex pairs beyond $T_{BKT}$.
It is well known that these vortices exhibit long-range interactions
which vary logarithmically with the distance between the vortices.
Vortices also occur in bulk Type-\Rmnum{2} superconductors in the
presence of an external magnetic field. Earlier experiments probing
vortices in bulk films through measurements of flux flow noise and
voltage noise have reported the presence of broad band noise (BBN)~\cite{Marley1995,Merithew1996}
as a function of driving current and magnetic field. The statistics
of noise in these systems is non-Gaussian due to the fact that most
of the noise arise from very few fluctuators. Noise measurements on
quasi-2D MgB$_{2}$ films show a possibility of thermally induced
vortex hopping through measurements of flux noise close to BKT transition~\cite{khare2008}.
\textcolor{black}{To the best of our knowledge there are no reports
on higher-order statistics of fluctuations looking into the possibility
of interacting vortices in two-dimensions.}

Higher-order fluctuations in resistivity has been established as an
useful tool to study the presence of long-range correlations in systems
undergoing both electronic and structural phase transitions. Experiments
on a wide variety of systems like doped silicon~\cite{Swastik2003},
metal nanowires~\cite{AmritaAPL}, shape memory alloys~\cite{Chandni_prl,ChandniActa},
disordered magnetic alloys~\cite{Petta1998} etc. have shown that
time dependent fluctuations (noise) in physical observables exhibit
a strong non-Gaussian behavior due to the presence of long-range correlations
originating from coulomb forces, internal strain fields or magnetic
interactions. The origin of non-Gaussian noise can be understood from
the following simple picture: In a system composed of many independent
fluctuators, the resultant noise has a Gaussian nature as expected
from the central limit theorem. However, the non-Gaussian component
increases when the correlation length becomes larger, for example
close to a critical phase transition, becoming maximum when the fluctuators
are correlated throughout the entire sample size. In reduced dimensions,
non-Gaussian fluctuations have been predicted for magnetization of
2D XY model which shows BKT transition as a function of temperature~\cite{Bramwell2001}.
However, there have been very few experiments probing the statistics
of fluctuations in systems undergoing BKT transition~\cite{Joubaud2008}.
In this letter we study the statistics of resistance fluctuations
to probe the nature of transition in thin film superconductors both
in 3D and quasi-2D limits. 

A characteristic feature of superconducting thin films is that when
the film thickness ($d$) is reduced below the coherence length ($\xi_{l}$),
the superconductivity crosses over to 2D limit characterized by BKT
transition~\cite{Mintu_PRL1,Vinokur2012}. The material system chosen
for our study is a conventional s-wave superconductor niobium nitride
(NbN). The Ginzburg-Landau coherence length ($\xi_{l}\sim6$~nm)
in these films was estimated from the measurement upper critical field
($H_{c2}$) as a function of temperature and using the relation $\xi_{l}(T)=\sqrt{\frac{\Phi_{0}}{2\pi H_{c2}(0)}}$,
where $H_{c2}(0)$ is the upper critical field extrapolated to $T=0$~K~\cite{Madhavithesis}.
NbN thin films exhibit a crossover from BCS to BKT type behavior,
with reduction in $d$ and are very good candidates to study the BKT
transition since ultra-thin films (down to $3$~nm) can be synthesized
with high degree of structural uniformity, while displaying moderately
large $T_{c}$. Thus they provide us an ideal platform to investigate
vortex correlations in low-dimensional superconductivity. 

The $T-d$ phase diagram of NbN films, showing the critical temperature
($T_{BKT}$ or $T_{BCS}$) vs normalized film thickness is shown in
Fig.~1a. Films with $d\sim\xi_{l}$, undergo BKT transition (black
line and symbol) while the expected mean field temperature is shown
by red line and symbol. The thicker/3D films ($d>\xi_{l}$) display
BCS type behavior and hence, are characterized by only one temperature
scale ($T_{BCS}$) whereas films in the quasi-2D limit ($d\lesssim\xi_{l}$),
which display BKT transition, are characterized by two temperature
scales ($T_{BKT}$ and $T_{BCS}$). The mean field critical temperature,
$T_{BCS}$, decreases as the film thickness is reduced due to the
increase in electron-electron coulomb interactions which compete with
the attractive interaction required for superconductivity~\cite{Finkelstein1994}.
The shaded region corresponds to the BKT fluctuation regime dominated
by interacting vortices.
\begin{figure}[tbh]
\begin{centering}
\includegraphics[width=0.48\textwidth]{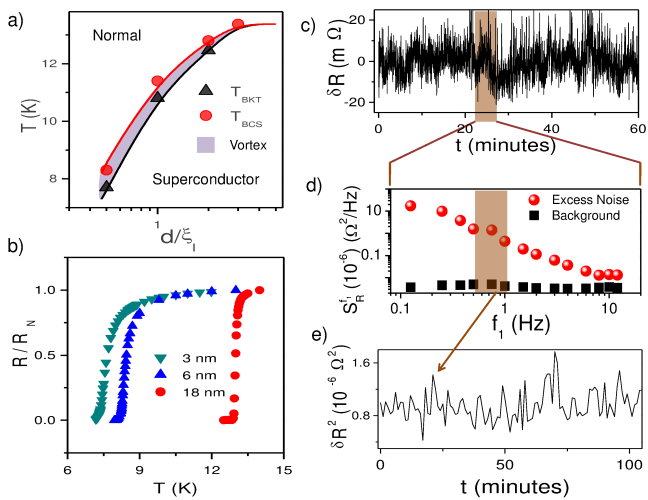}
\par\end{centering}

\caption{(a) Phase diagram of thin films of NbN as a function of normalized
film thickness, where $\xi_{l}$ is the superconducting coherence
length (extracted from Table.~1 in Ref.~\cite{Mintu_PRL1}). Shaded
region denotes presence of vortices due to BKT transition in films
in quasi-2D limit. The red and black lines are guide to eye representing
the variation in $T_{BCS}$and $T_{BKT}$ respectively. (b) Normalized
resistance for films of different thickness as a function temperature
where $R_{N}$ represents the normal state resistance. (c) Time series
of resistance fluctuations at $7.3$~K for $d=3$~nm. (d) Power
Spectral Density (PSD) for a given time window. (e) Fluctuations in
noise power in an octave spanning $0.5-1$~Hz across different time
window.}
\end{figure}

The samples used in our experiments were grown on lattice matched
MgO substrate using RF magnetron sputtering (details refer~\cite{ChokalingamPRB,MadhaviNbN,Mintu_PRL1}).
Contacts (Cr/Au) were defined by wire masking for four-probe measurements.
The normalized temperature dependence of resistance is shown for all
films in Fig.~1b. The width of transition from normal state to superconducting
state is $\approx1.1$~K for thicker films and increases to $\approx1.8$~K
when the thickness is reduced below $6$~nm. It is important to note
that $T_{c}$ depends very strongly on the microstructural details,
and in spite of very similar deposition conditions, variations in
the microstructral details, particularly grain size, can make $T_{c}$
vary over nearly $30$~\%. This also indicates a limitation of standard
time-averaged transport to characterize the nature of ultra-thin superconductivity
without ambiguity.

Our experimental procedure consists of measuring slow time-dependent
fluctuations in the sample resistance at different temperatures for
zero magnetic field. Typical time traces of the fluctuations for a
$3$~nm thick NbN film are shown in Fig.~1c and Fig.~2a. The details
of the noise measurement technique are available in Ref.~\cite{ChandniActa}.
As the temperature is reduced, the fluctuations increase in magnitude
(Fig.~2a). The normalized power spectral density (PSD) for different
temperatures as shown in Fig.~2b, exhibits $1/f$$^{\alpha}$ type
behavior. The PSD increases in magnitude as the temperature is lowered.
We find that $\alpha\sim2-4$ at low temperatures, which eventually
converges to $\sim1$ at higher temperatures restoring the $1/f$
nature of noise. We find similar behavior in $\alpha$ even for thicker
samples. The rather high values of $\alpha$ might possibly arise
due to the presence of vortices / inhomogeneous percolative transport~\cite{ShoboVorMatter}.
We believe that further experiments are required to understand this
behavior of $\alpha$. 
\begin{figure}[tbh]
\begin{centering}
\includegraphics[bb=0bp 0bp 630bp 540bp,width=0.52\textwidth]{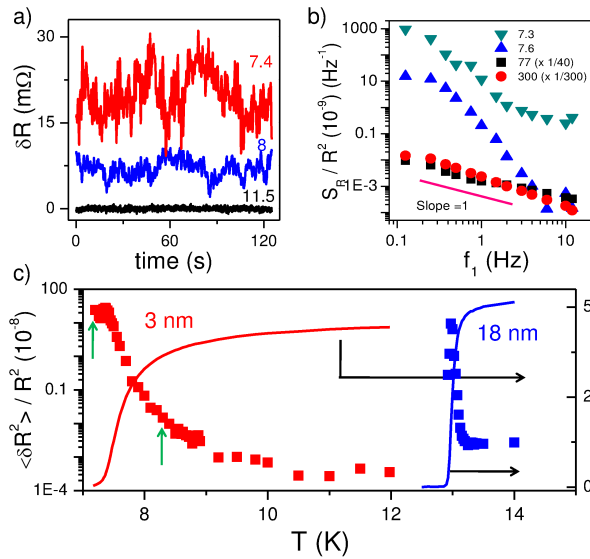}
\par\end{centering}

\caption{(a) Time traces showing the resistance fluctuations at different temperatures
with for $d=3$~nm. The traces are displaced by $10$~m$\Omega$
for clarity. The mean value of resistance at $7.4$, $8$ and $11.5$~K
are $5.3$, $33$ and $44.2$~$\Omega$ respectively (b) PSD at different
temperatures for $d=3$~nm. Scaling factors are indicated in the
legend for $T=77$~K and $T=300$~K. (c) Normalized variance of
noise ($\langle\delta R^{2}\rangle/R^{2}$) (filled squares) and resistance
($R$) (line) at different temperatures for $d=3$~nm (red) and $d=18$~nm
(blue). The green arrows indicate the temperatures at which second
spectrum is shown in Fig.~3a. }
\end{figure}
 To explore the temperature dependence of noise close to the superconducting
transition, we integrate the PSD over our measurement bandwidth ($\approx15$~Hz)
to obtain the normalized variance which is shown along with resistance
as a function of temperature in Fig.~2c. We find that the normalized
variance shows a divergent behavior at low temperatures for all samples
irrespective of thickness. This behavior is best understood from the
theory of percolation. It is well known that for $T>T_{BKT}$ ($T>T_{BCS}$
for $d\geq18$~nm) thin film superconductors exhibit fluctuations
in the superconducting order parameter which leads to formation of
cooper pairs for a very short time. These fluctuations in the superconducting
order parameter leads to the formation of small superconducting islands
in the films, giving rise to a percolative network of metallic and
superconducting regions. As the temperature is reduced, the superconducting
phase grows. The noise in these systems arise from two possible scenarios
: a) resistance fluctuations in regions of normal metal and b) fluctuations
in the number of superconducting regions/islands. It is well known
from the theory of percolation that noise grows till the film becomes
fully superconducting~\cite{Kiss1993,KissPRL1993,KIssIEEE1994,Testa1988}.
The noise follows the relation $\langle\delta R^{2}\rangle/R^{2}\varpropto R^{-l_{rs}}$
where the percolation exponent $l_{rs}\sim1\pm0.4$ in our experiments.
The influence of temperature fluctuations ($\ll0.5$~mK) was carefully
evaluated and eliminated by varying the control parameters and observing
its effect on the nature of the noise power spectrum.

To probe the presence of correlations, we measured higher-order statistics
of noise using second spectrum of the fluctuations. The second spectrum
is a sensitive technique to estimate the presence of non-Gaussian
component (NGC) in the resistance fluctuations~\cite{Restle1985,Seidler1996,Chandni_prl}.
The experimental scheme employed to measure the second spectrum is
shown in Fig.~1c-1e. The time series shown in Fig.~1c is divided
into successive windows of $2$~minutes each. The power spectral
density is calculated for each such window (Fig.~1d) and it is integrated
over a chosen octave to obtain the noise power which is plotted as
a function of time in Fig.~1e. In effect, the second spectrum is
the Fourier transform of the four-point correlation of $\delta R\left(t\right)$
and is given by 
\[
S_{R}^{f_{1}}\left(f_{2}\right)=\int_{0}^{\infty}\langle\delta\mathit{R}^{2}(t)\delta R^{2}(t+\tau)\rangle\mbox{\ensuremath{\cos}}\left(2\pi f_{2}\tau\right)\mathrm{\, d}\tau
\]

where $f_{1}$ is the center frequency of the chosen octave and $f_{2}$
corresponds to the spectral frequencies. $S_{R}^{f_{1}}\left(f_{2}\right)$
represents the ``spectral wandering'' or fluctuations in the noise
power in a chosen frequency band $\left(f_{L},f_{H}\right)$ (Fig.~1e).
Specifically, we have chosen an octave spanning $0.5-1$~Hz where
the excess noise is considerably higher than the background to minimize
the effects of signal corruption by Gaussian background. The second
spectrum of noise appears to display a rather weak dependence of frequency
but shows an increase in overall magnitude with lowering of temperature
(Fig.~3a). A convenient way to represent the second spectrum is to
plot the normalized variance of second spectrum $\sigma^{(2)}=$$\nicefrac{\int_{0}^{f_{H}-f_{L}}S_{R}^{f_{1}}(f_{2})df_{2}}{\left[\int_{f_{L}}^{f_{H}}S_{R}(f)df\right]^{2}}$,
which is a measure of normalized fourth-order moment in noise, as
a function of temperature. As shown in Fig.~3b, $\sigma^{(2)}$ decreases
monotonically from $\approx9$ at $7.2$~K to a baseline value of
$\approx4$ near $7.5$~K. Comparison with the temperature dependence
of resistance reveals that the enhanced $\sigma^{(2)}$ marks the
onset of the normal state, which decays to a temperature-independent
baseline value within about $10$~\% of the normal state resistance
($R_{N}$). Importantly, $\sigma^{(2)}$ is reduced to the baseline
value by $T\approx7.6$~K, whereas the percolation dynamics of the
resistance fluctuations is observable well up to $\approx10$~K (Fig.~2c).
This implies that the increase in $\sigma^{(2)}$ is due to an independent
process, and unlikely to be due to percolative fluctuations that dominate
the measured noise (first spectrum).
\begin{figure}[tbh]
\begin{centering}
\includegraphics[width=0.4\textwidth]{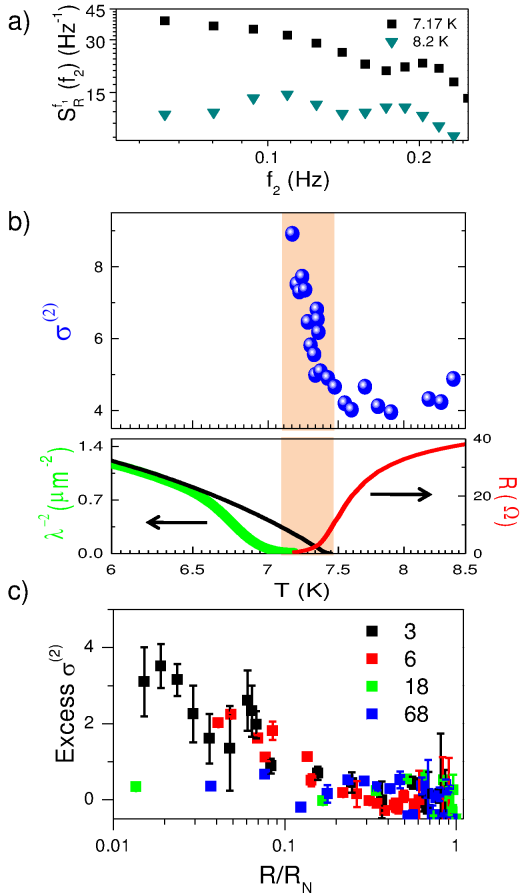}
\par\end{centering}

\caption{(a) Second spectrum at different temperatures indicated by green arrows
in Fig.~2c in an octave spanning $(0.5-1)$~Hz . (b) Temperature
dependence of normalized variance of second spectrum $\sigma^{\left(2\right)}$(blue
line), superfluid density $\lambda^{-2}\left(T\right)$ (green line)
and sample resistance $R$ (red line). BCS fit for $\lambda^{-2}\left(T\right)$
is shown by black line. Black arrows point to the y-axis. (c) Excess
$\sigma^{\left(2\right)}$as a function of normalized resistance $(R/R_{N})$
for samples of different thickness. The error bars were calculated
from measurements of $\sigma^{(2)}$ over $60$ time windows.}
\end{figure}

To correlate the measured $T$-dependence of $\sigma^{(2)}$ with
other physical properties of the NbN thin films at the superconductor-normal
transition, we have subsequently measured the magnetic penetration
depth ($\lambda$) in the same sample as a function of $T$~\cite{Kamlapure2010}.
In such measurements $\lambda^{-2}$ is proportional to the superfluid
density, whose deviation from the mean field BCS expectation (black
line in Fig.~3b) has recently been interpreted as the BKT transition
in ultra-thin NbN films~\cite{Mintu_PRL1}. In the present sample
(Fig.~3b), the maximum in $\sigma^{(2)}$ occurs at the same temperature,
suggested as the BKT transition temperature $T_{BKT}$, where the
superfluid density ($\lambda^{-2}$) drops to zero. Strikingly, $\sigma^{(2)}$
approaches the baseline value at the theoretically computed mean-field
superconducting transition temperature ($T_{BCS}$), suggesting a
correlated kinetics of the charge carrying species between $T_{BKT}$
and $T_{BCS}$ (the shaded region in Fig.~3b). 

To get a further understanding of the underlying mechanism, we have
repeated the measurements for NbN films with different thickness.
We subtract the baseline value from $\sigma^{\left(2\right)}$ in
each case and plot the excess value as a function of normalized resistance
$R/R_{N}$ for each sample (Fig.~3c). We find that the representation
in terms of $R/R_{N}$ allows better comparison as the samples display
a wide range of $T_{C}$. The dependence of $\sigma^{(2)}$ on $R/R_{N}$
for thinner films ($d\sim\xi_{l}$) was found to be very different
from the thicker ones ($d\geq18$~nm). Most ($>80\%$) quasi-2D samples
($d\leq6$~nm) show the remarkable increase in $\sigma^{\left(2\right)}$
close to the superconducting transition signifying the presence of
long-range correlations, whereas the bulk films display essentially
no excess $\sigma^{\left(2\right)}$ which is indicative of Gaussian
type fluctuations as expected in case of Ginzburg-Landau (GL) fluctuations
for a BCS superconductor. The thickness dependent behavior of $\sigma^{(2)}$
has two key implication: First, since all samples, irrespective of
thickness, show percolative kinetics near the superconducting transition,
we can eliminate the enhancement of $\sigma^{(2)}$ to arise from
such kinetics. Secondly, appearance of excess $\sigma^{(2)}$ in the
regime dominated by vortex fluctuation (Fig.~3b and schematics of
Fig.~1a) strongly suggests the excess $\sigma^{(2)}$ to arise from
long-range correlations among the vortices themselves.

Another phenomenon which can give rise to NGC in noise is Dynamical
Current Redistribution (DCR) which occurs due to large local resistivity
fluctuations and strong transport inhomogeneities~\cite{seidlerDCR}.
In our samples, $\delta R/R\ll1$ and the noise mechanism (percolation)
is the same for both thin and thicker films. Had the origin of NGC
been due to DCR, then all the films must have displayed strong NGC
in noise which is not the case in our measurements. 
\begin{figure}[tbh]
\begin{centering}
\includegraphics[width=0.48\textwidth]{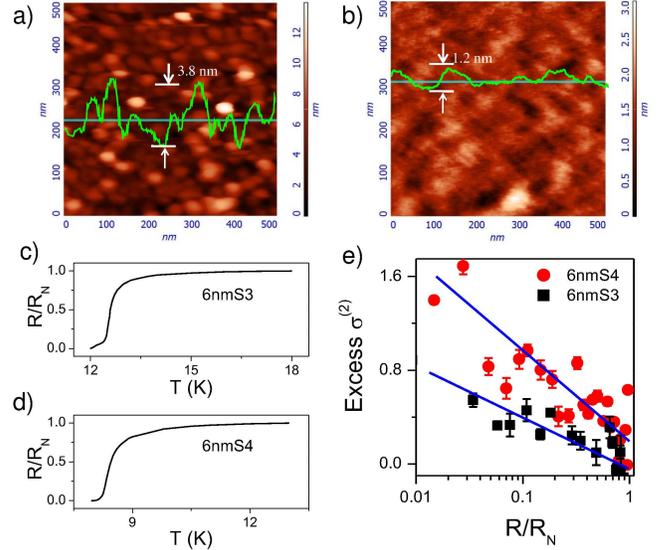}
\par\end{centering}

\caption{(a), (b) AFM images of samples ($d=6$~nm) 6nmS3 and 6nmS4 respectively
with the insets showing a line scan (green line) indicating the roughness
of respective samples. (c), (d) RT characteristics of both the samples.
(e) Excess $\sigma^{\left(2\right)}$ for both the samples. Solid
blue lines are guide to eye.}

\end{figure}

Finally, to explore the sensitivity of non-Gaussian fluctuations to
the structural details of the films, we have carried out measurements
on two different samples of same nominal thickness 6~nm, but different
structural morphology. We used an AFM to characterize the roughness
of the samples that were prepared at different times and deposition
conditions. One of the sample (6nmS3) shows an average roughness of
$~$$1.65$~nm, whereas the relatively smoother sample (6nmS4) has
a much lower average roughness ($0.31$~nm) (Fig.~4a,b). We observe
a kink like feature (around $12.2$~K) in the resistance vs temperature
data (Fig.~4c) for 6nmS3, possibly as a result of the surface roughness
and strong thickness variation in the film. A clear distinction is
observed in the second spectrum of noise (Fig.~4e) with the NGC becoming
weaker as the surface roughness increases. These results can be viewed
in the context of vortices in disordered/granular superconductors.
Granular thin films are modeled as disordered array of Josephson junctions
which can lead to a double transition~\cite{JohnsonPRB1998,Bergk_NJP2011}.
Moreover, the pinning of vortices due to random potential fluctuations/defects
is known to limit the long-range interactions among the vortices~\cite{GiamarchiPRB95,GiamarchiPRB97}.
Further experiments on samples with controlled levels of disorder
will be of great interest in evaluating the robustness of NGC in noise. 

Thus our experiments underline a new method to identify the BKT fluctuation
regime in ultra-thin superconducting films. The characteristics temperature
scales $T_{BKT}$ and $T_{BCS}$ emerge as the range over which the
NGC is non-zero, that could be useful to probe other low-dimensional
superconductors as well.
\begin{acknowledgments}
We acknowledge the Department of Science $\&$ Technology (DST) and
Department of Atomic Energy (DAE), Govt. of India for funding the
work.
\end{acknowledgments}
\bibliographystyle{apsrev}
\bibliography{NbN}

\end{document}